\documentclass[showpacs,preprintnumbers,amsmath,amssymb, preprint]{revtex4}

\usepackage{graphicx}
\usepackage{dcolumn}
\usepackage{bm}
\usepackage{color}

\begin{document}

\preprint{Shafran, Tuning Near-Field Interactions}

\title{Tuning Near-Field Interactions Using the Bright and Dark States of a Quantum Dot}

\author{Eyal Shafran}
\author{Benjamin D. Mangum}
\altaffiliation{Current address:  Soft Matter Nanotechnology and Advanced Spectroscopy
Team, Los Alamos National Laboratory}
\author{Jordan M. Gerton}
\email{jgerton@physics.utah.edu}
\affiliation{Department of Physics and Astronomy, University of Utah, 115 S. 1400 E., Salt Lake City, UT 84112}

\date{\today}

\begin{abstract}
We demonstrate that the cycling between internal states of quantum dots during fluorescence blinking can be used to tune the near-field coupling with a sharp tip.
In particular, the balance between tip-induced field enhancement and energy transfer depends explicitly on the intrinsic quantum yield of the quantum dot.
Our measurements show that for internal states with low quantum yield, energy transfer is strongly suppressed in favor of field enhancement, and explicitly demonstrate that suppressed blinking of quantum dots near metal surfaces is due to fast energy transfer.
\end{abstract}

\maketitle

The near-field interaction between a dipole emitter and a nanoscale structure is an intriguing problem that is fundamentally interesting and also important for a number of applications, including tip-enhanced microscopy and surface-enhanced spectroscopy. 
The field enhancement generated near sharp edges and tips via the electrostatic lightning-rod effect \cite{Gerton2004, Protasenko2004,Protasenko2004NL, Yoskovitz2008, Mangum2009} and/or geometry-dependent resonance effects such as surface plasmon polaritons \cite{Taminiau2008, Muhlschlegel2005, Farahani2005, Novotny2007}, leads to an increase in an emitter's photoexcitation rate for separation distances below $\sim$10 nm.
On the other hand, the nanostructure can also quench a fluorophore's emission by providing external non-radiative relaxation channels to which the emitter can couple directly \cite{Issa2007, Anger2006, Kuhn2006, Carminati2006}, and can additionally modify the local density of optical states, as in the Purcell effect \cite{Purcell1946}.
The net optical signal can thus be quite convoluted and difficult to interpret, except when one effect is dominant, or when the field enhancement can be reliably calculated, as is the case for metal nanospheres \cite{Anger2006, Kuhn2006, Carminati2006}.
For arbitrary metallic nanostructures, neither quenching nor enhancement can be neglected, and deconvolution is difficult.  

The intrinsic quantum yield ($q_0$) of an emitter plays a central role in moderating its interaction with the environment.
In particular, a large value of $q_0$ indicates that internal non-radiative relaxation processes are slow relative to radiative emission, so even weak external coupling (quenching) will noticeably decrease the fluorescence signal.
On the other hand, a small value of $q_0$ indicates fast internal relaxation, making the fluorophore relatively insensitive to external coupling, and thus more sensitive to field enhancement.  
While some experiments have shown that larger signal enhancement factors are obtained for lower values of $q_0$ \cite{Protasenko2004, Frey2009}, no rigorous study has ever been performed.
Here we demonstrate how different $q_0$ values, corresponding to different internal relaxation rates of an emitter, can be used to modify its interaction with a sharp tip by tuning the balance between quenching and enhancement. 
Large variations in the net fluorescence signal are observed, including a clear contrast reversal for gold tips. 
A simple analytical model is used to deconvolute the enhancement and quenching portions of the signal, revealing that gold tips exhibit both strong quenching and enhancement, while silicon tips exhibit nearly as strong enhancement but very weak quenching \cite{Gerton2004, Protasenko2004,Protasenko2004NL, Yoskovitz2008, Mangum2009}, and carbon nanotube (CNT) tips exhibit very strong quenching and no enhancement \cite{Shafran2010}.
Finally, our measurements demonstrate explicitly that reduced blinking from quantum emitters adsorbed onto metal surfaces is caused by fast energy transfer, in agreement with recent observations of quantum dots on metal surfaces \cite{Yuichi2007}, nanoparticle films \cite{Ma2010} and ITO \cite{Jin2010}.

\begin{figure}[b]
	\centering
		\includegraphics[width = 8.6 cm]{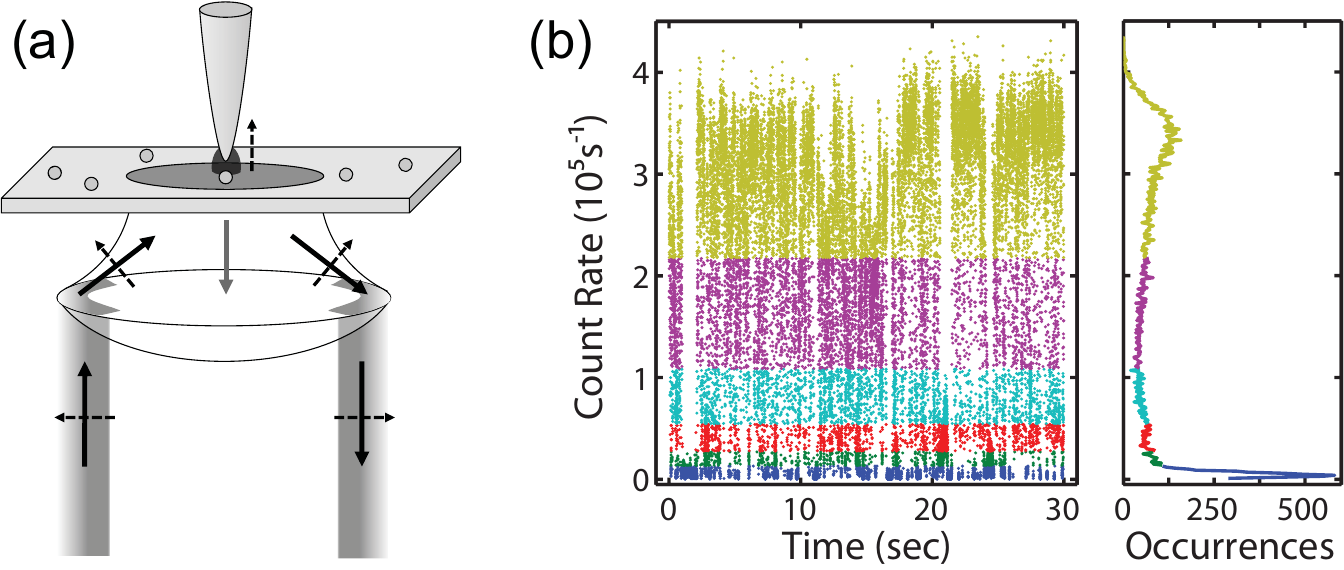}
	\caption{Experimental scheme.  (a) Excitation light is incident on the sample above the critical angle, creating an evanescent field above the sample with polarization along the axis of the tip. Fluorescence data are collected as the tip oscillates vertically above an isolated QD.  (b) Photoluminescence trajectory from a single QD. The different colors illustrate the threshold levels in Fig. \ref{fig2}(a,b).}
	\label{fig:Schematic}
\end{figure}

To investigate how the quantum yield of an emitter affects its near-field interactions, we utilize the well-known phenomenon of fluorescence intermittency (blinking) in single semiconductor nanocrystal quantum dots (QDs) \cite{Efros1997}.  An example of a QD photoluminescence trajectory is shown in Fig. \ref{fig:Schematic}(b).
Current theories suggest that the temporal fluctuations in emission intensity result from the transition between a neutral state for which $q_0 \sim 1$ and a charged state for which $q_0 < 1$ . 
Charged states are thought to originate from strongly localized trapping of an exciton's electron or hole at the periphery of the QD core; charged states with intermediate values of $q_0$ are sometimes called ``gray'' states.
Furthermore, single-photon counting experiments have revealed a correlation between the decrease in emission intensity and a decrease in the fluorescence lifetime \cite{Schlegel2002, Fisher2004, Zhang2006, Montiel2008}, leading to the conclusion that a single QD can have many different surface traps with potentially different Auger recombination rates \cite{Zhang2006}.
These studies in combination with extinction coefficient measurements showing similar absorption cross-sections for low and high emissive states \cite{Kukura2009} suggest that the temporal fluctuations of the emission rate from individual QDs are due to variations in $q_0$ corresponding to a relatively large number ($\sim$10) of trap states. 
We leverage these dynamic fluctuations in individual QDs to tune the near-field interactions via the value of $q_0$.  
In particular, we use a novel photon counting technique \cite{Mangum2009} to capture interaction curves as a function of both $q_0$ and the spatial separation between a sharp atomic force microscope (AFM) tip and an individual QD.

As an AFM tip approaches the QD from above, the fluorescence signal will increase or decrease due to the various mechanisms described above when the tip enters the interaction range; beyond this range, the fluorescence is unaffected by the tip.
Normalizing the fluorescence signal to this far-field rate gives,
\begin{equation}
\label{signal}
S_{\rm norm}(z) = \frac{q(z)}{{q}_0}\kappa(z) =  \frac{\Gamma_r+\Gamma^\prime_r(z)}{\Gamma_0 + \Gamma^\prime(z)}\frac{\kappa(z)}{q_0},
\end{equation}
where $q(z)$ is the effective quantum yield including modifications induced by the tip at height $z$ above the QD, $\Gamma_0 = \Gamma_{\rm r} + \Gamma_{\rm nr}$ is the intrinsic fluorescence rate including radiative ($\Gamma_{\rm r}$) and non-radiative ($\Gamma_{\rm nr}$) relaxation channels, and $\Gamma^\prime(z) = \Gamma^\prime_{\rm r}(z) + \Gamma_{\rm et}(z)$ is the {\em tip-induced} relaxation rate.  
In this notation, $\Gamma^\prime_{\rm r}(z)$ can be negative corresponding to a tip-induced suppression of the radiative rate and $\Gamma_{\rm et}(z)$ is the non-radiative energy transfer rate from QD to tip.
Finally, $\kappa(z) = I(z)/I_0$ is the far-field normalized excitation intensity at the QD, which includes tip-induced near-field effects.

Equation (\ref{signal}) can be re-parameterized as:
\begin{equation}
\label{signal2}
S_{\rm norm}(z) = \frac{\alpha}{1+ q_0 \beta}, 
\end{equation}
where $\alpha(z) \equiv \kappa(z) \left[ 1+ \Gamma^\prime_r(z) / \Gamma_r \right]$ characterizes tip-induced changes to the local excitation intensity and the radiative rate, and $\beta(z) \equiv \Gamma^\prime(z) / \Gamma_r$ characterizes the strength of the near-field coupling. 
For fixed $q_0$, $S_{\rm norm}$ can only increase if $\alpha$ increases, while an increase in $\beta$ leads to a decrease in $S_{\rm norm}$.
Thus, the shape of a vertical approach curve measurement, $S_{\rm norm}(z)$, for a particular value of $q_0$ reflects the dynamic balance between $\alpha$ and $\beta$.
The values of $\alpha$ and $\beta$ are extracted from measurements of $S_{\rm norm}$ for different values of $q_0$, as demonstrated below.

For light polarized along the $z$-axis ($p$-polarized), the intensity at the tip apex should be enhanced due to the lightning-rod effect and/or surface plasmons. 
Furthermore, at short tip-sample distances $z$, the radiative rate of the fluorophore can be enhanced (i.e. $ \Gamma^\prime_r>0$) due to the Purcell effect. 
These two mechanisms will increase $\alpha$ leading to signal enhancement. 
On the other hand, superposition of direct and tip-scattered excitation light at the QD can lead to a decrease in $\kappa(z)$ on a wavelength scale \cite{dHili2003, Mangum2009, Kuhn2006}; $\alpha$, however, is typically dominated by near-field enhancement for $z < 50$ nm. 
The larger the product, $q_0\beta$, the stronger the reduction in $S_{\rm norm}$, so clearly a decrease in $q_0$ makes $S_{\rm norm}$ less sensitive to tip-sample coupling (quenching), and thus more sensitive to signal enhancement.
Furthermore, fluctuations in $q_0$ are amplified when $\beta$ is large (i.e., strong near-field coupling), leading to dramatic changes in $S_{\rm norm}$.

Data were obtained using a  tip-enhanced fluorescence microscope (TEFM), which utilizes an AFM sitting atop a custom optical setup \cite{Mangum2009}.
A continuous wave helium-neon laser ($\lambda = 543$ nm) is used as the excitation source.
A small wedge of supercritical rays are allowed into the back aperture of a microscope objective (NA=1.4) such that QDs are illuminated with an evanescent field of intensity $\sim$350 W/cm$^2$, Fig. \ref{fig:Schematic}(a).
A half-wave retarder allows for easy manipulation of the polarization state. 
The emitted photons are collected by the same objective and are focused onto an avalanche photodiode. 
The tip is aligned into the center of the focused illumination spot and the sample is raster scanned.
The AFM is operated in tapping mode with typical oscillation frequencies of 60-80 kHz and peak-to-peak amplitudes of 200-250 nm, depending on the specific probe. 
Several silicon, gold-coated, and home-made carbon nanotube (CNT) \cite{Hafner2001,Mu2008} AFM probes were used for the measurements described below. 
The sample consisted of elongated ($4 \, {\rm nm} \times 9 \, {\rm  nm}$) CdSe/ZnS QDs emitting at 605 nm, diluted in toluene and dried onto a glass coverslip. 
All data were taken at room temperature.

To extract vertical approach-curve measurements, $S_{\rm norm}(z)$, the lateral sample scan is halted when the tip is directly above a QD for the duration of a measurement ($\sim$30 s), and every detected fluorescence photon is then timestamped. 
Each signal photon is then correlated with the instantaneous height of the tip above the QD \cite{Gerton2004, Mangum2009}. 
The photoluminescence trajectory from an individual QD is constructed using 1 ms time bins (Fig. \ref{fig:Schematic}(b)), and each photon is assigned a far-field count-rate value corresponding to its particular bin. 
Three important parameters are thus encoded with each signal photon:  time of emission, tip height at the time of emission, and far-field count-rate value at that point in the fluorescence trajectory. 
The background signal is calibrated by moving the sample to an area with no QDs, and is subsequently subtracted carefully from the data.

\begin{figure}[tb]
	\centering
		\includegraphics[width = 8.6 cm]{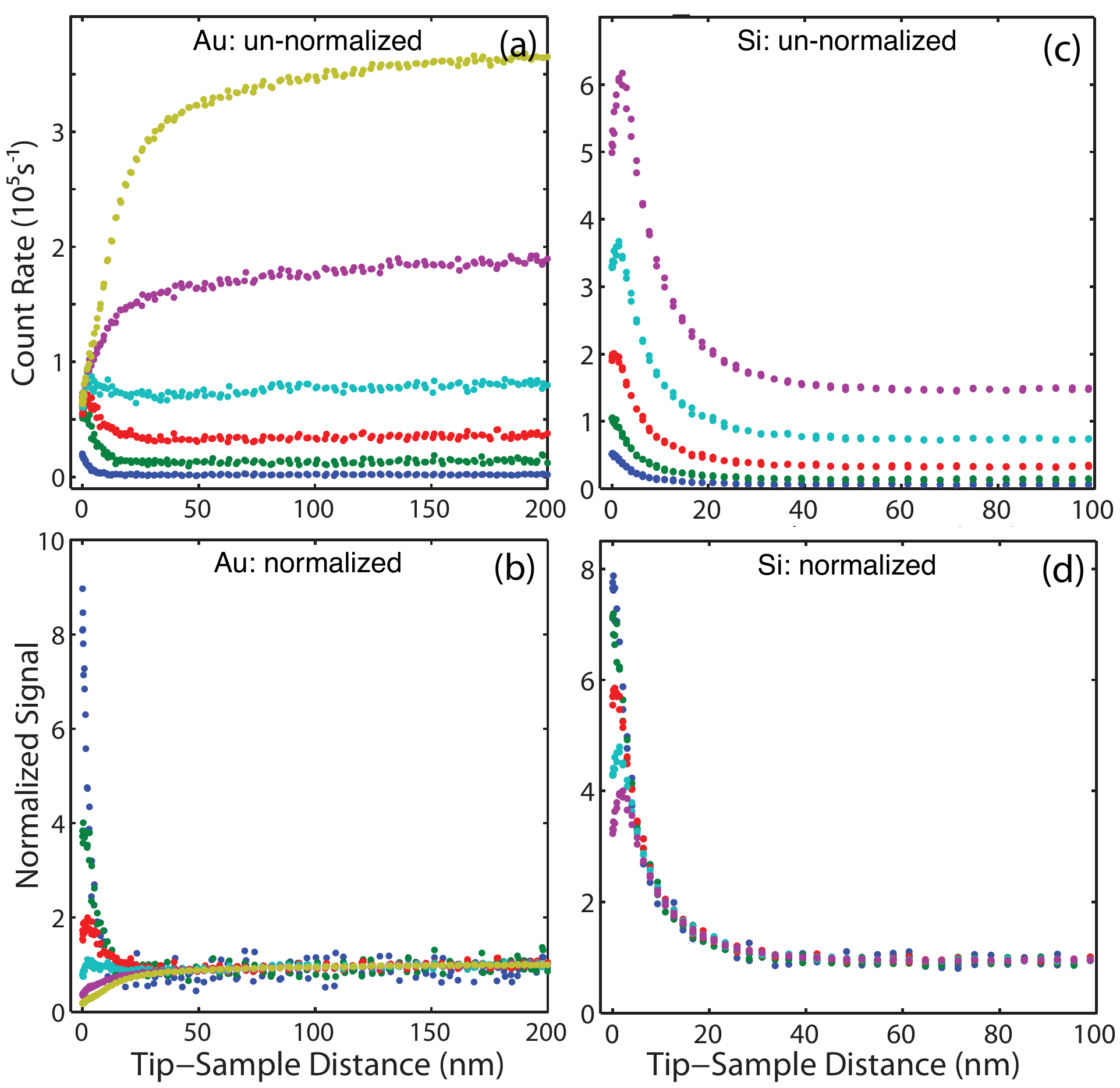}
	\caption{Vertical approach curves for different fluorescence intensity thresholds.  Un-normalized (a, c) and normalized (b, d) fluorescence signals as a function of the tip height above a QD for gold-coated (a, b) and silicon (c, d) probes.  The color-coding in (a, b) corresponds to the threshold values shown in Fig. \ref{fig:Schematic}(b).  For each tip, the various approach curves were obtained from a single measurement as the QD cycled dynamically through different quantum yield states.}
	\label{fig2}
\end{figure}

The data acquisition technique described above enables tip-sample approach curves to be reconstructed for different emission intensity thresholds, i.e., different values of $q_0$.   
Figure \ref{fig2} shows the un-normalized (a, c) and normalized (b, d) QD fluorescence signal as a function of the tip-sample distance for a gold-coated (a, b) and silicon (c, d) tip. 
For each tip, the various approach curves shown were extracted from a single 30-s measurement on a single QD:  the photon data were first separated into various emission intensity thresholds, as illustrated in  Fig. \ref{fig:Schematic}(b), and the $z$-dependent fluorescence signal was then reconstructed for each. 
The various threshold ranges differ successively by a factor of two, which emphasizes the differences between the corresponding values of $q_0$. 

\begin{figure}[b]
	\centering
		\includegraphics[width = 8.6 cm]{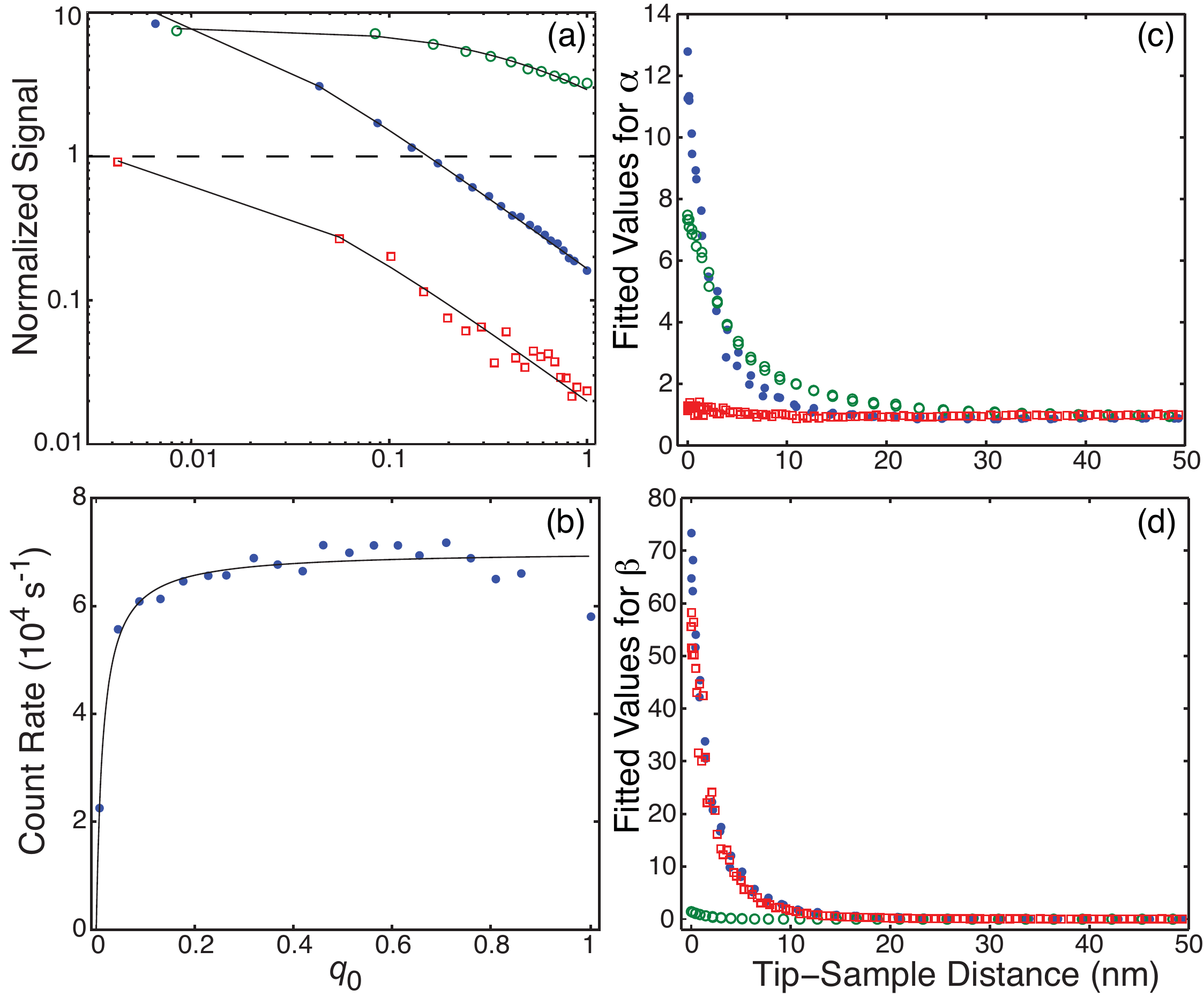}
	\caption{Separation of enhancement and quenching.  (a) Normalized fluorescence signal at $z=0$ and corresponding fit to Eq. \ref{signal2} for gold-coated (blue circles), silicon (green circles), and CNT (red squares) tips. The dashed horizontal line denotes the crossover from enhancement to quenching. (b) Un-normalized fluorescence signal at $z=0$ for the gold-coated tip in (a); the solid line is the best fit to Eq. \ref{signal3}. Panels (a) and (c) show the fitted values for $\alpha$ and $\beta$ for different tip-sample distances.}
	\label{fig3}
\end{figure}

For the gold-coated tip, Fig. \ref{fig2}(a, b), the bright states of the QD are strongly quenched as fast energy transfer to the tip outcompetes relatively slow intrinsic relaxation.
In contrast, the darkest states are strongly enhanced as fast intrinsic relaxation outcompetes energy transfer, yielding more sensitivity to the enhanced field at the tip apex. 
The ratio of signal enhancement factors for the darkest state compared to the brightest one is $\sim$$9/0.16 = 56$ at $z=0$, as shown in Figs. \ref{fig2}(b) and \ref{fig3}(a) (note that a signal enhancement factor below unity indicates quenching). 
Thus, strong near-field coupling (large $\beta$) for gold tips amplifies changes in $q_0$ according to Eq. \ref{signal2}, leading to a clear contrast reversal for bright and dark states. 
On the other hand, weak coupling between the silicon tip and QD imparts poor sensitivity to changes in $q_0$, as seen in Figs. \ref{fig2}(d) and \ref{fig3}(a). 
In this case, field enhancement is dominant for all values of $q_0$, although the competition between enhancement and quenching does cause a minor decrease in signal at the smallest tip-sample separation distances.
As $q_0$ becomes smaller, intrinsic relaxation of the QD outcompetes energy transfer at progressively smaller tip heights, until finally the signal increases monotonically as the tip approaches the QD.

Figure \ref{fig3}(a) plots $S_{\rm norm}$ at $z = 0$ as a function of $q_0$ for gold-coated, silicon, and CNT tips, where we have assumed that $q_0 = 1$ for the brightest state of a particular photoluminescence trajectory \cite{Fisher2004, Brokmann2004}.
For these data, the highest far-field threshold corresponded to $\geq 70\%$ of the maximum count, in agreement with previous observations \cite{Fisher2004}.
The remainder of each trajectory was divided linearly into as many distinct ranges as possible, so as to allow for sufficient signal-to-noise ratio in each.
The different threshold ranges arise from different electron- or hole-trap states through which the QD cycles, each with a slightly different value of $q_0$.
If the number of trap states is sufficiently large ($\gtrsim$10), the system can be regarded as continuous and the results of the analysis will not be sensitive to the particular number of threshold ranges, $N$. 
This was verified by varying $N$ between $\sim$12 and 30 for each measurement.
The data for each trajectory in Fig. \ref{fig3}(a) were fit to Eq. \ref{signal2}; the fits (solid lines) for all three tips are excellent across more than 100-fold variation in $q_0$.
The dashed line represents the crossover from signal enhancement to quenching:  the near-field signal is dominated by enhancement for silicon tips, quenching for CNT tips, and can be tuned via $q_0$ to either enhancement or quenching for gold-coated tips.

The analysis was repeated at each value of tip-QD separation, and the fitting parameters $\alpha$ and $\beta$ were extracted, as shown in Fig. \ref{fig3}(c, d). 
The maximum value for $\alpha$ is larger for gold-coated tips compared to silicon, reflecting stronger field enhancement at this wavelength.
Without the aid of simulations or a geometry-dependent analytical model, it is not possible to determine this difference unambiguously with any other method, since gold-coated tips also quench the signal strongly:  $\beta_{\rm Au} / \beta_{\rm Si} \sim 37$ at $z=0$. 
The CNT tip induces no significant enhancement at any value of $q_0$ or $z$ ($\alpha_{\rm CNT} \sim 1$), but does quench the signal strongly indicating efficient energy transfer between the QD and CNT at short separations \cite{Shafran2010}. 

Importantly, these measurements expose the mechanism responsible for suppressed blinking of quantum emitters adsorbed onto metal surfaces \cite{Yuichi2007,Ma2010,Jin2010}.
Fig. \ref{fig3}(b) plots the un-normalized signal at $z=0$ as a function of $q_0$, showing that the measured fluorescence rate is approximately constant down to $q_0 \sim 0.1$.
Thus, as a QD blinks, it samples states with different $q_0$ values, and if it were adsorbed onto a metal surface, corresponding to $z \sim 0$, these different states would yield similar fluorescence rates.
This is the origin of suppressed blinking.

The suppressed blinking is also predicted by the simple parameterization of the signal given above.
Multiplying Eq. \ref{signal2} by $q_0$ yields an expression that is proportional to the un-normalized fluorescence signal, $S(z)$:
\begin{equation}
\label{signal3}
S(z) \propto q_0 \, S_{\rm norm}(z) = \frac{q_0 \alpha}{1+q_0 \beta}.
\end{equation}
Thus, for large values of $\beta$, as for a gold-coated tip at $z=0$, the product $q_0 \beta \gg 1$ and until $q_0 \rightarrow 0$, the un-normalized signal will be independent of $q_0$:  $S(z) \propto \alpha / \beta$.
Importantly, suppressed blinking is thus expected to occur for any geometry or material for which $\beta$ is large.
The solid curve in Fig. \ref{fig3}(b) is the best fit line corresponding to Eq. \ref{signal3}, with $\alpha = 12.8$ and $\beta = 73.4$, as extracted from Fig. \ref{fig3}(c, d), and the proportionality factor as the only fitting parameter.
When $q_0 \beta \gg 1$, the fluorescence rate (relative to the far-field value) converges to $\alpha / \beta$, which depends on the probe geometry and material.
For gold-coated tips, $\alpha / \beta \sim 0.17$ at $z=0$, so the fluorescence signal is nearly six-fold smaller than the maximum far-field value, and the lack of blinking is clearly discernable.
For CNT tips, $\alpha / \beta \sim 0.02$ so the fluorescence signal is suppressed by nearly a factor of 50, and it is difficult to detect the lack of blinking above the noise level.

In conclusion, the dynamic fluctuations in quantum yield that occur during QD blinking tune the balance between tip-induced enhancement and quenching, and make it possible to separate their contributions to the net fluorescence signal.
Our measurements show that the near-field signal is dominated by enhancement for silicon tips, quenching for CNT tips, and can be tuned via $q_0$ to either enhancement or quenching for gold-coated tips.
In addition, gold-coated tips strongly suppress fluorescence blinking, in agreement with previous observations of QDs adsorbed onto metal surfaces.
Our measurements demonstrate explicitly that the lack of blinking is the result of strong near-field coupling (large $\beta$) between the tip and emitter, which results from efficient energy transfer from the emitter to the tip ($\Gamma_{\rm et} \gg \Gamma_r$).
In principle, the blinking should also be suppressed if the radiative rate becomes large, $\Gamma^\prime_r \gg \Gamma_r$, which might occur in an optical cavity, photonic crystal, or near a plasmonic nanoantenna.

This work was supported in part by an NSF CAREER Award (DBI-0845193) and a Cottrell Scholar Award from the Research Corporation for Science Advancement.


\end{document}